# Toward Phonon-Limited Transport in Two-Dimensional Electronics by Oxygen-Free Fabrication

Subhrajit Mukherjee,[1,2] Shuhua Wang,[3] Dasari Venkatakrishnarao,[1] Yaoju Tarn,[1] Teymour Talha-Dean,[1,4] Rainer Lee,[1,2] Ivan A. Verzhbitskiy,[1,2] Ding Huang,[1,2] Abhishek Mishra,[1,2] John Wellington John,[1,2] Sarthak Das,[1,2] Fabio Bussoloti,[1,2] Thathsara D. Maddumapatabandi,[1] Yee Wen Teh,[3] Yee Sin Ang*,[3] Kuan Eng Johnson Goh*,[1,2,5,6] and Chit Siong Lau*,[1,2,3]

[1]*Institute of Materials Research and Engineering (IMRE), Agency for Science, Technology and Research (A*STAR), 2 Fusionopolis Way, Innovis #08-03, Singapore 138634, Republic of Singapore*
[2]*Quantum Innovation Centre (Q.InC), Agency for Science Technology and Research (A*STAR), 2 Fusionopolis Way, Innovis #08-03, Singapore 138634, Republic of Singapore*
[3]*Science, Mathematics and Technology, Singapore University of Technology and Design, 8 Somapah Road, 487372, Singapore*
[4]*Department of Physics and Astronomy, Queen Mary University of London, London E1 4NS, U.K.*
[5]*Department of Physics, National University of Singapore, 2 Science Drive 3, Singapore 117551, Singapore*
[6]*Division of Physics and Applied Physics, School of Physical and Mathematical Sciences, Nanyang Technological University, 50 Nanyang Avenue, Singapore 639798, Singapore*

Email: aaron_lau@imre.a-star.edu.sg; kejgoh@yahoo.com; yeesin_ang@sutd.edu.sg

**Future electronics require aggressive scaling of channel material thickness while maintaining device performance. Two-dimensional (2D) semiconductors are promising candidates, but despite over two decades of research, experimental performance still lags theoretical expectations. Here, we develop an oxygen-free approach to push the electrical transport of 2D field-effect transistors toward the theoretical phonon-limited intrinsic mobility. We achieve record carrier mobilities of 91 (132) $cm^2V^{-1}s^{-1}$ for mono- (bi-) layer $MoS_2$ transistors on $SiO_2$ substrate. Statistics from over 60 devices confirm that oxygen-free fabrication enhances key figures of merit by more than an order of magnitude. While previous studies suggest that 2D transition metal dichalcogenides such as $MoS_2$ and $WS_2$ are stable in air, we show that short-term ambient exposure can degrade their device performance through irreversible oxygen chemisorption. This study emphasizes the criticality of avoiding oxygen exposure, offering guidance for device manufacturing for fundamental research and practical applications of 2D materials.**

    Silicon complementary metal-oxide-semiconductor (CMOS) technology has been pivotal in the progress of modern electronics. Advances in nanofabrication techniques have enabled continued scaling of CMOS transistor technology.[1] Meeting the growing demand for cheaper, faster, and more energy-efficient electronics will require the ultimate scaling of the semiconductor channel thickness toward the sub-nanometre regime. However, achieving this with any three-dimensional semiconductor is challenging due to the increased surface roughness scattering from dangling bonds that severely degrades carrier mobility.[2] 2D transition metal dichalcogenide (TMD) semiconductors with intrinsic atomic-scale thickness (<1 nm) could be a solution.[3–9] Their van der Waals (vdW) layered nature provides dangling-bond-free surfaces such that high carrier mobilities are maintained even in monolayer devices. This has positioned 2D TMDs on the technology roadmap of industry, including companies

such as the Taiwan Semiconductor Manufacturing Corporation (TSMC), Samsung, Intel, and IMEC.[10] However, despite promising early device prototypes, 2D TMD transistor performance falls short of their theoretical expectations.[11–13] Realizing their full potential will depend on developing high-quality material growth, controlled doping, dielectric integration, and contact engineering methods.[4,7,9,14] Encouragingly, there has been considerable progress in these areas. Chemical vapor deposition growth can produce highly crystalline materials with fewer defects and impurities to rival exfoliated material quality.[15–17] The vdW interfaces formed between TMDs and In[18,19], Bi[20], Pt,[21] and Sb[22,23] three-dimensional (3D) metals, as well as Y-doping[24] of source-drain contact regions, have led to ultralow contact resistances. Surface[25], substitutional[26], and sub-stoichiometric oxide[27,28] doping have increased sheet conductance and on-state currents. Dielectric and substrate engineering to quench impurity and phonon scattering have boosted TMD carrier mobilities.[29–33] These advances have narrowed the experiment-theory device performance gap and showed that progress depends on tailored solutions that consider the unique nature of 2D materials: their extreme geometry.

2D geometry with extreme surface-to-volume ratios heightens the sensitivity of 2D materials to their environment.[34–38] A monolayer TMD comprises a three-atom crystal lattice where the layer of transition metal atoms is sandwiched between two layers of chalcogens. The top layer of chalcogen atoms forms a surface that can interact with ambient molecular absorbates. There are four possible interaction scenarios for molecular adsorbates on TMD surfaces (illustrated in Figure 1a).[35] These are generally divided into weakly interacting physisorption and strongly interacting chemisorption. Physisorption of a molecular adsorbate occurs on a (i) sulfur atom or (ii) a sulfur vacancy ($V_S$), while chemisorption may exhibit either (iii) non-dissociative or (iv) dissociative coupling to a $V_S$. For dissociative chemisorption, the molecule dissociates into individual atoms which substitutionally replaces a $V_S$ site. This substitution causes a disruptive and often permanent stoichiometry change, e.g., chemical oxidation. Ambient oxidation is an undesired material degradation of TMDs. It is well documented, as changes in stoichiometry are easily detected by experimental techniques like X-ray photoemission spectroscopy (XPS) (Extended Data Figure 1) or Raman and photoluminescence spectroscopy.[38–43] In contrast, physisorption (i, ii) is often considered a reversible process where the smaller interaction energies allow adsorbate removal through thermal annealing or vacuum cycling.[44,45] Consequently, TMDs such as $MoS_2$ and $WS_2$ are often considered ambient stable as many studies have confirmed their short-term resistance against (iv) dissociative chemical oxidation. However, the impact and reversibility of non-dissociative chemisorption (iii) remain elusive. Although some theoretical studies[35] have hinted at its significance, this process remains poorly understood due to challenges in experimental validation, especially at the device level, where ambient exposure is unavoidable in standard cleanroom fabrication.

In this work, we study the impact of ambient molecular exposure on the device performance of 2D TMD transistors. Using a purpose-built facility, we can complete the entire fabrication of 2D devices in an inert argon environment. We found that minimizing oxygen exposure throughout the fabrication process results in more than an order of magnitude enhancement across key transistor figures of merit, such as carrier mobility, sheet conductance, and carrier density. This enables us to push our device carrier mobility toward the theoretical limit arising from intrinsic phonon scattering.[11–13,46] Recognizing molecular oxygen as a critical factor limiting device performance sets a significant milestone. This marks a qualitative step in bridging the theoretical-experimental discrepancy and guides the manufacture of high-performance 2D devices for fundamental research and practical applications.

To understand the impact and likelihood of physisorption and non-dissociative chemisorption (hereafter referred to as chemisorption in the rest of the manuscript for simplicity), we first perform adsorption energy ($E_{ad}$) and charge transfer ($\Delta q$) analysis using density functional theory (DFT) simulations (details in Methods, Extended Data Figs 2-5, and SI Figs S1-S7). The simulations evaluate the interactions of common ambient molecules $O_2$, $N_2$, and $H_2O$ with monolayer $MoS_2$ and $WS_2$. Using a 4×4×1 supercell, we consider both pristine (no $V_S$) and defective (single $V_S$) monolayers. Such a supercell size corresponds to a defect density of ~7.1×10$^{13}$ cm$^{-2}$, comparable to expected experimental values.[47]

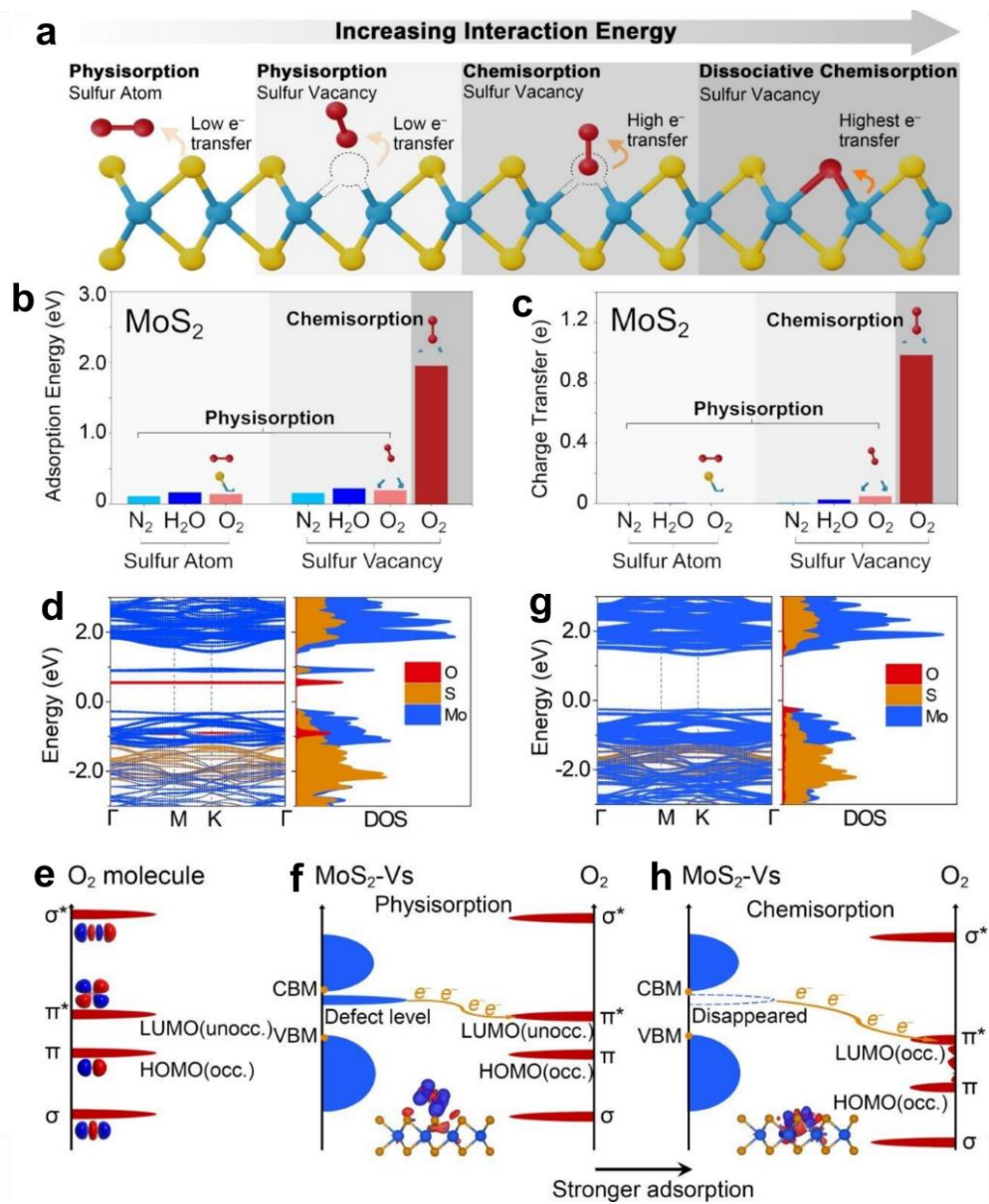

**Figure 1: Molecular adsorption on monolayer TMDs.** (a) Schematic of possible diatomic molecular adsorbate interaction scenarios with a monolayer TMD. The dotted outlines represent a sulfur vacancy $V_S$. (b) Adsorption energy of $O_2$, $N_2$, and $H_2O$ on the pristine surface and sulfur vacancy sites of a monolayer $MoS_2$, and resultant (c) electron transfer from TMDs to molecular adsorbates. Electronic band structure and the projected density of states (PDOS) of $MoS_2$ with sulfur vacancies $V_S$ during $O_2$

(d) physisorption and (g) chemisorption. (e) Molecular orbital diagram of the $O_2$ molecule showing the highest occupied molecular orbital (HOMO) and lowest unoccupied molecular orbital (LUMO). Schematic of the electron transfer process and local electronic structure during $O_2$ physisorption (f) and chemisorption (h) on defective $MoS_2$. The embedded plot shows the charge density difference, where blue regions represent electron accumulation and red regions represent electron depletion. Electrons transfer from TMD to the LUMO of $O_2$ and accumulate on the $O_2$ molecule in the shape of π orbitals.

We calculate the $E_{ad}$ of $O_2$, $N_2$, and $H_2O$ for $MoS_2$ (Figure 1b) and $WS_2$ (Extended Data Figure 5a). In pristine TMDs, all studied molecules show weak physisorption with characteristic energies $E_{ad}$ ~114 – 176 meV. In contrast, sulfur vacancies in TMDs support the chemisorption of $O_2$ to the $V_S$ site, while interactions with $N_2$ and $H_2O$ remain limited to physisorption. Notably, $O_2$ chemisorption is significantly stronger than physisorption, with $E_{ad}$ values almost an order of magnitude larger, $E_{ad}$ ~1.95 - 2.34 eV. Consequently, this strong interaction between $O_2$ and $V_S$, coupled with the large electronegativity of molecular oxygen, should result in substantial charge transfer at the chemisorption site. Indeed, our Bader charge analysis confirms this behavior (Figure 1c and Extended Data Figure 5b). $O_2$ chemisorption leads to considerably more charge transfer with $\Delta q = 0.91e$ for $MoS_2$ (Figure 1c) and $1.08e$ for $WS_2$ (Extended Data Figure 5b) compared to physisorption, where $\Delta q < 0.049e$ for $O_2$, $N_2$, and $H_2O$. As further validation, calculations performed using different-sized supercells and defect densities show similar trends for $E_{ad}$ and $\Delta q$ (SI Figure S1 and Table-S1). The calculated band structure shows that $V_S$ on the surface of TMDs act as electron donors and induce localized defect states in the band gap. When $O_2$ molecules are physisorbed at the $V_S$ site, the lowest unoccupied molecular orbitals (LUMO) of $O_2$ are slightly below the defect levels, suggesting potential electron transfer from the TMD to the $O_2$ molecules (Figure 1d-f). In contrast, the chemisorption of $O_2$ fills the $V_S$ sites with oxygen atoms, leading to the disappearance of the defect level. Electron transfer results in the complete occupation and delocalization of the $O_2$ LUMO (Figure 1g and h). These simulation results suggest that $O_2$ exposure affects TMD device performance. Even for ambient stable TMDs with low defect densities, the large $E_{ad}$ of $O_2$ chemisorption can lead to unwanted and irreversible p-doping. Uncontrolled doping from ambient molecular adsorbates at the source/drain and channel regions leads to detrimental effects such as larger parasitic contact resistance, undesired threshold voltage shift, and lowered sheet conductance.

However, experimentally assessing the impact on devices from strongly chemisorbed $O_2$ is challenging. So far, studies have measured the effects on 2D TMD devices from weakly physisorbed molecules and shown that it is reversible through vacuum and thermal cycling.[44,45] Evaluating the impact from potentially irreversible chemisorbed molecules will require minimal ambient exposure, even during material synthesis and device fabrication. Currently, 2D TMDs are commonly synthesized through mechanical exfoliation or chemical vapor deposition (CVD). Exfoliation is usually conducted in standard laboratory or cleanroom environments. It can also be performed relatively quickly in controlled environments, such as an inert argon-filled glovebox with minimal $O_2$ and $H_2O$ levels. However, for CVD-grown TMDs, limiting ambient exposure is difficult due to the technical challenge of integrating large-growth furnace tubes with gloveboxes. More significant concerns are the materials processing and device fabrication steps. CVD films are typically wet transferred from growth to device substrates. Also, device fabrication will require cleanroom lithography and thin film deposition processes, which must now be in controlled inert conditions.

Therefore, to experimentally evaluate the impact of molecular chemisorption on TMD devices, we designed a solution for TMD device fabrication with minimal ambient exposure. Our approach is based on purpose-built gloveboxes integrated with nanolithography, thin film deposition, and solvent processing capabilities (Extended Data Figure 6), with typical $O_2$ and $H_2O$ levels maintained at ~1 ppm. For this study, we fabricated sets of $MoS_2$ and $WS_2$ back-gated FETs on 290 nm $SiO_2/Si^{++}$, utilizing both exfoliated and CVD materials. All devices fabricated in our glovebox facility are hereafter referred to as 'inert.' We fabricated equivalent device sets in ambient conditions as a control, hereafter referred to as 'ambient.' In total, 38 $MoS_2$ and 24 $WS_2$ devices were measured for statistical analysis of FET metrics, including threshold voltage, sheet conductance, mobility, hysteresis, and interfacial trap densities.

Our theoretical results suggest that $O_2$ chemisorption leads to substantial charge transfer doping. We can determine the degree of charge doping by comparing the threshold voltages $V_{th}$ of ambient and inert devices. Room temperature transfer curves of representative exfoliated monolayer $MoS_2$ and $WS_2$ inert (blue) and ambient (red) devices are shown in Figures 2a and 2b, where we observe typical *n*-type semiconducting behavior. Consistent with our theoretical expectations, ambient devices exhibit substantial *p*-doping due to electron transfer from TMDs to the molecular adsorbates. This *p*-doping is characterized by significant shifts in $V_{th}$ toward positive gate voltages with $\Delta V_{th}$ = 70 V ($MoS_2$) and 35 V ($WS_2$) corresponding to an electron doping of ~$5.1 \times 10^{12}$ cm$^{-2}$ ($MoS_2$) and ~$3.3 \times 10^{12}$ cm$^{-2}$ ($WS_2$). Figure 2c ($MoS_2$) and 2d ($WS_2$) show statistics summarizing $V_{th}$ distributions. We observe significant doping for $MoS_2$ where mean $\Delta V_{th}$ = 90 V (exfoliated) and 65 (CVD) corresponding to an electron *n*-doping of ~$6.5 \times 10^{12}$ cm$^{-2}$ (exfoliated) and ~$4.7 \times 10^{12}$ cm$^{-2}$ (CVD). For $WS_2$, we find more modest doping with mean $\Delta V_{th}$ = 19 V (exfoliated) and 3 V (CVD) corresponding to an electron doping of ~$1.3 \times 10^{12}$ cm$^{-2}$ (exfoliated) and ~$2.2 \times 10^{11}$ cm$^{-2}$ (CVD).

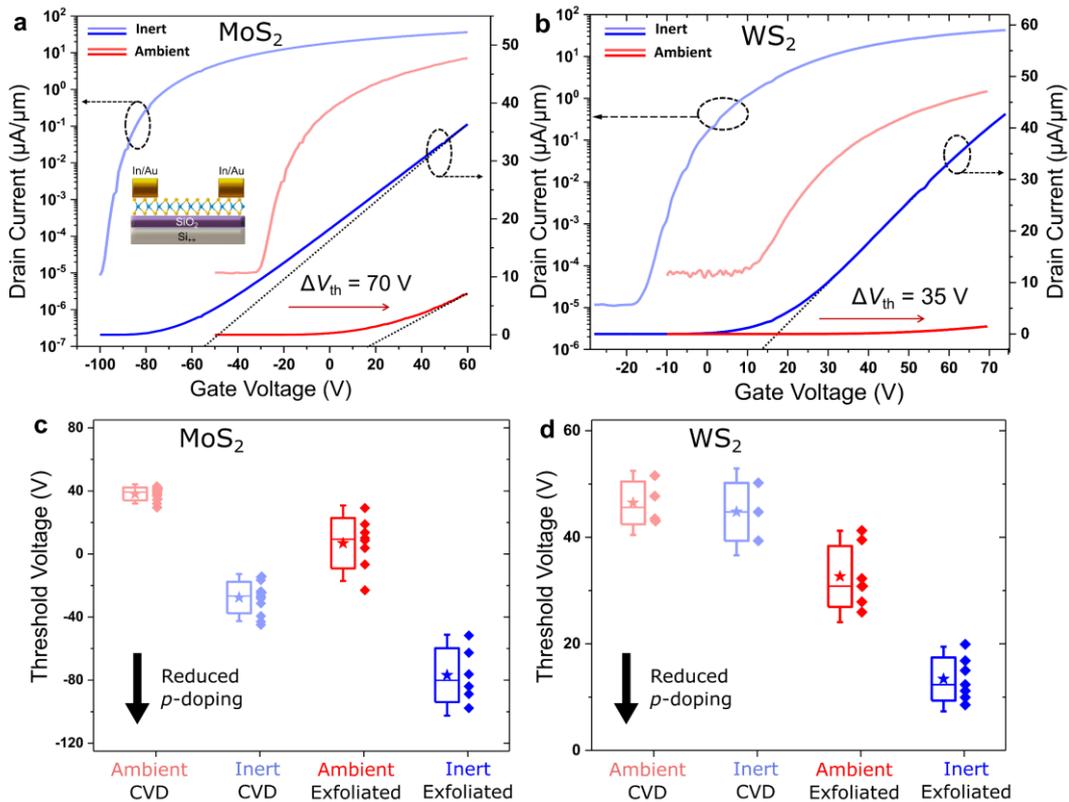

**Figure 2. Charge doping from oxygen.** Transfer characteristics of a monolayer (a) $MoS_2$ and (b) $WS_2$ back-gate field-effect transistor at a drain voltage $V_D$ =1 V. The device schematic is shown as an inset. The device channel lengths are 1 μm. The dotted lines represent the linear extrapolation to obtain threshold voltages ($V_{th}$). Boxplots present statistical distributions of the extracted threshold voltages for (c) $MoS_2$ and (d) $WS_2$ devices. Significant negative shifts in threshold voltages are observed for inert devices compared to their ambient counterpart, which indicate reduced $O_2$-driven $p$-doping.

The observed difference in doping levels between $MoS_2$ and $WS_2$ is surprising. Our DFT calculations suggest negligible differences in Δ$q$, and defect densities are also expected to be lower in $MoS_2$ crystals with better growth quality, which should present fewer $V_S$ sites for $O_2$ chemisorption. A possible explanation is that more defective $WS_2$ crystals tend to undergo dissociative oxidation from defect pairing.[35,41] The kinetics for dissociative oxidation likely differ from chemisorption and may be material-dependent. Furthermore, differences in overall stoichiometry, purity, and crystallinity are expected between $MoS_2$ and $WS_2$ crystals, suggesting more complex $O_2$ interaction mechanisms. However, the difference in doping levels observed between CVD and exfoliated materials is expected. Exfoliated devices represent the lowest defect density and ambient exposure levels where 2D flakes are prepared inside the gloveboxes. For commercially purchased CVD TMDs, the post-growth packaging and shipment process is beyond our control. CVD devices thus represent higher defect densities and a 'medium' ambient exposure level, translating to more significant $p$-doping.

Minimizing ambient exposure to prevent unwanted $p$-doping enables our devices to reach high carrier densities up to ~1.2 × $10^{13}$ $cm^{-2}$ ($MoS_2$) and ~5.2 × $10^{12}$ $cm^{-2}$ ($WS_2$). For context, our $MoS_2$ carrier densities approach degenerate doping levels[48] and are comparable with intentional surface transfer or solid-state doping techniques (~1-2 × $10^{13}$ $cm^{-2}$).[25,27,48] As a result, inert devices can attain more than an order of magnitude larger on-state current $I_{ON}$ and current on/off ratios than their ambient counterparts (Figure 2a, b). Our inert monolayer $MoS_2$ FET reached $I_{ON}$ ~280 μA/μm (SI Figure S8), exceeding even ultra-short channel (~29 nm) FETs developed by IMEC (250 μA/μm).[49] Current on/off ratios for inert devices (>$10^6$) exceed ambient devices (~$10^4$-$10^5$) by around an order of magnitude. Such significant $n$-doping was previously only achieved by intentional doping techniques. However, such doping introduces foreign impurities that lead to impurity-related Coulomb scattering or band structure changes impacting electrical transport, lowering carrier mobilities and sheet conductance. Our DFT calculations confirm that the chemisorption of $O_2$ molecules can lead to band structure changes that increase effective mass and lower carrier mobility (Extended Data Figure 9). In contrast, such effects are not expected for our approach, which minimizes ambient adsorbates to preserve the intrinsic cleanliness of 2D TMDs.

To investigate whether minimizing ambient adsorbates translates into superior electrical transport, we extract the carrier mobility and sheet conductance from two-terminal room-temperature transport measurements of 38 $MoS_2$ (Figure 3a) and 24 $WS_2$ (Figure 3b) few-layer and monolayer devices (details in Methods). Clear improvements are observed for inert (blue) compared to ambient (red) devices made from both exfoliated (diamond markers) and CVD (circle markers) materials. For exfoliated (CVD) $MoS_2$, we find that the mean mobility increases by a factor of ~3 (3) to 86.5±16.7 $cm^2$/Vs (25.7±5.8 $cm^2$/Vs), while the mean sheet conductance increases by a factor of ~9 (6) to 138.8±25.5 μS/□ (16.5±3.6 μS/□). Similar improvements are observed for exfoliated (CVD) $WS_2$, with mean mobility increasing by a factor of ~10 (26) to 35.9±1.8 $cm^2$/Vs (18.5±1.4 $cm^2$/Vs), and the mean sheet conductance increases by a factor of ~26 (10) to 23.6±1.9 µS/□ (7.1±0.8 µS/□) (details are in Extended Figure 10). At room (low) temperature, mobilities are typically phonon- (impurity-) scattering limited. Our measurements at 80 K show that mobility increases by a factor of ~4 (3) to 470.8±0.2 cm$^2$/Vs (106.3±0.3 cm$^2$/Vs) for exfoliated $MoS_2$ ($WS_2$) devices. We measure similarly high mobility improvement by a factor of ~3 (3) for monolayer CVD $MoS_2$ ($WS_2$) devices. In contrast, the exfoliated ambient $MoS_2$ device shows only a modest 19% improvement, suggesting that higher impurity levels from ambient exposure limit the low-temperature carrier mobility. All analyzed devices exhibit linear output behavior at both room and low temperatures, ruling out poor contacts as the cause of the different mobility trends (Extended Data Figure 11) between ambient and inert devices. Transfer length method measurements corroborate our device contact quality. We find that minimizing ambient exposure is also beneficial for contact quality. Inert devices show lower contact resistance $R_c$ = 10±2 kΩµm compared to ambient devices where $R_c$ = 15±50 kΩµm, (Extended Data Figure 12). This lower $R_c$ can be understood by the decreased p-doping of contact regions which decreases Schottky barrier heights.[66] The performance advantage of inert devices also extends to other figures of merit where we find consistent improvements in gate hysteresis and interface trap densities (Extended Data Figure 13).

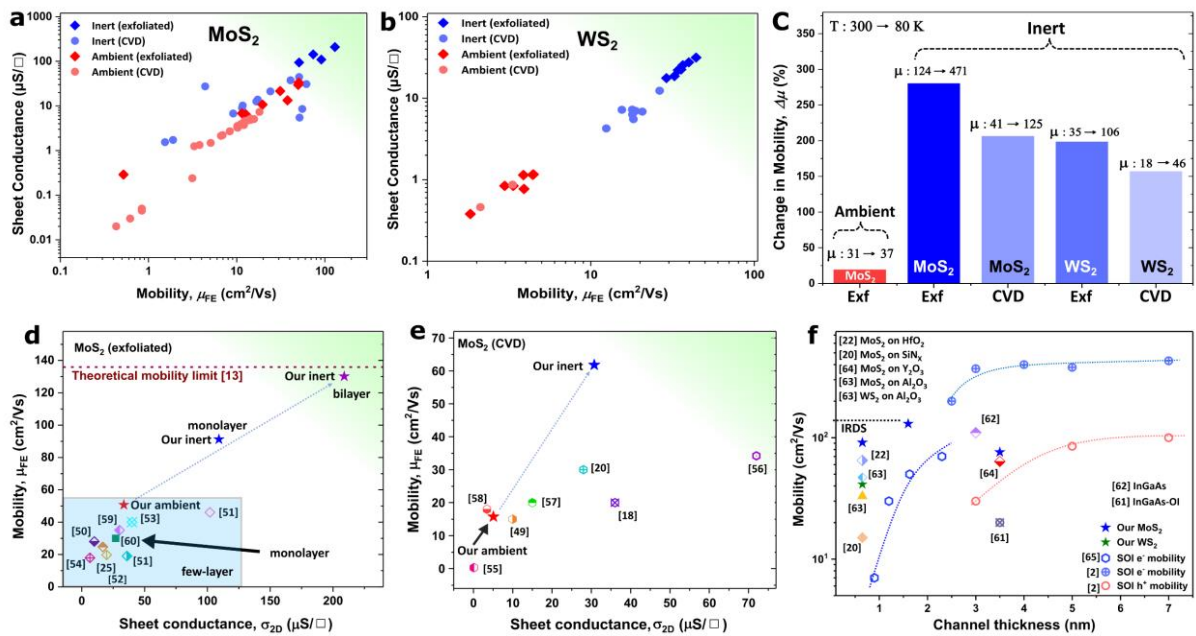

Distribution of field-effect mobility ($\mu_{FE}$) vs sheet conductance ($\sigma_{2D}$) for (a) $MoS_2$ and (b) $WS_2$ field-effect transistors (FETs) fabricated in inert (blue) and ambient (red) conditions. Mean mobility values show ∼2 orders of magnitude increase for inert devices compared to ambient devices. Similarly, sheet conductance is enhanced by ∼3 orders for $MoS_2$ and ∼2 orders for $WS_2$. (c) When the temperature is decreased from 300 K to 80 K, significant enhancement is observed for inert devices compared to ambient devices, showing higher impurity-limited mobility for inert devices. Benchmarking FET performance against the existing literature for (d) exfoliated $MoS_2$ and (e) CVD $MoS_2$. The red dashed line represents the theoretical limit of the room-temperature, phonon-limited mobility in 2D $MoS_2$. We benchmark the sheet conductance and back-gated field-effect mobility of FETs with comparable geometries on $SiO_2$. The data used for refs are provided as a table in the Supporting Information.[18,20,25,49–60] (f) We also benchmark mobility as a function of FET channel thickness. We compare our devices with other state-of-the-art 2D FETs, traditional ultra-thin body (UTB) silicon PMOS and NMOS, III-V 2DEG MOSFETs, and IRDS roadmap targets.[2,20,22,61–65]

Next, we benchmarked our device performance with other devices of similar architecture as reported in the literature. We summarize sheet conductance and mobility values of exfoliated 2D $MoS_2$ (Figure 3d), CVD monolayer (Figure 3e) $MoS_2$, and $WS_2$ (Extended data Figure 14) devices. Our exfoliated mono- and bi-layer $MoS_2$ FETs demonstrate record room-temperature mobility of ~91 $cm^2V^{-1}s^{-1}$ and ~132 $cm^2V^{-1}s^{-1}$ respectively, more than three times higher than previous reports of comparable device geometries. Likewise, we measure a record high mobility of 36 $cm^2V^{-1}s^{-1}$ for $WS_2$ FET. The comparatively modest improvement for $WS_2$ is likely due to crystal quality limitations such that transport is still defect-limited, whereas $MoS_2$ growth processes are more mature. Notably, our $MoS_2$ mobility marks a significant advance toward the theoretical expectations for phonon-limited transport in 2D TMDs (indicated by the horizontal red dashed line in Figure 3d). We have found that minimizing ambient exposure to molecular oxygen leads to superior devices. Notably, our fabrication strategy is also generally compatible with most reported contact, dielectric, and interface engineering techniques that can further improve 2D TMD device performance.

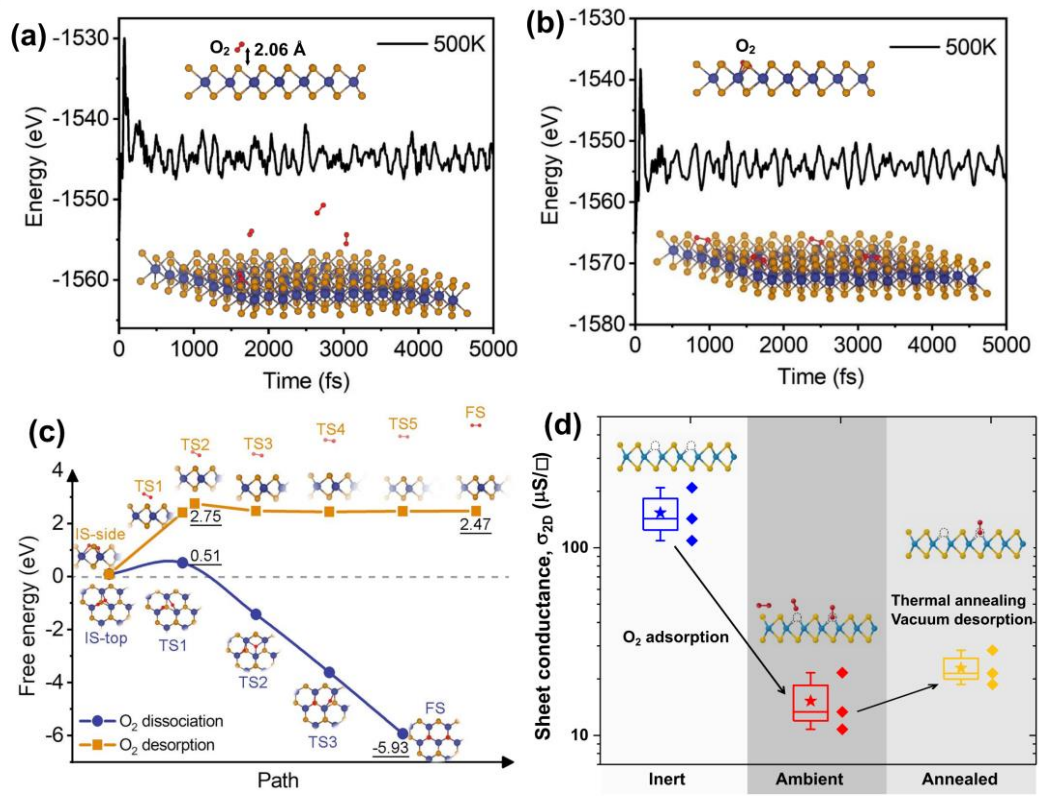

**Figure 4: Reversibility of $O_2$ chemisorption.** Ab initio molecular dynamics (AIMD) simulation of (a) physisorbed and (b) chemisorbed $O_2$ at 500 K. The simulations indicate that ~75% of physisorbed $O_2$ molecules desorb at 500 K, while no chemisorbed $O_2$ molecules desorb. (c) Gibbs free energy profile of the initial state (IS), transition state (TS), and final state (FS) of $O_2$ dissociation and desorption on $MoS_2$ with adjacent double sulfur vacancies. The calculations show that desorption becomes energetically unfavorable once the $O_2$ molecules are chemisorbed. Instead, the dissociation of $O_2$ molecules is the more energy-favorable pathway. (d) Sheet conductance of $MoS_2$ FETs inert and ambient devices (calculated from Extended Figure 15). The ambient devices were subsequently annealed at 473 K in a 10% $H_2$/Ar forming gas environment for two hours. Annealing partially recovers the conductance attributed to the desorption of physisorbed $O_2$. However, chemisorbed $O_2$ molecules restrict its complete recovery.

These measurements and benchmarking highlight the significant impact of $O_2$ adsorption at the device level. Next, we investigate the reversibility of adsorption processes. Based on our DFT calculated $E_{ad}$ (Figure 1b), weakly interacting physisorption should be reversible through simple annealing or vacuum cycling treatments. However, the large $E_{ad}$ of $O_2$ chemisorption poses a significant barrier toward molecular desorption and will require considerable thermal energies. While 2D TMDs can remain structurally intact up to 500 °C,[67] devices may not tolerate similarly high temperatures. In integrated circuit manufacturing, back-end-of-line (BEOL) compatibility is an essential consideration with a temperature limit of ~400 °C.[5] Thermal effects such as strain and atomic diffusion at the contacts, dielectrics, and interfaces can degrade device performance and reliability at even lower temperatures below 300 °C.[22] With these considerations, we perform ab initio molecular dynamics (AIMD) simulation to understand the desorption process at elevated temperatures. We simulate $O_2$ molecules physisorbed and chemisorbed to TMD surfaces with double $V_S$ (Figure 4a, b). At a temperature of 500 K, 75% (0%) of physisorbed (chemisorbed) $O_2$ molecules can desorb from the TMD surface. Importantly, our simulations (Figure 4c) suggest that chemisorbed $O_2$ molecules are more likely to undergo dissociative oxidation than desorb from the TMD surface. This is consistent with our previous work, where XPS measurements confirmed the tendency for dissociative oxidation in defective 2D $WS_2$ with double $V_s$.[41] Next, we experimentally assess the reversibility of device performance after thermal annealing at 200 °C for 2h. Figure 4d shows the box plots of sheet conductance measured for inert, ambient, and annealed devices. Inert devices show the highest mean sheet conductance of 153.7±29.3 µS/□ while ambient devices show the lowest mean sheet conductance of 15.2±3.3 µS/□. After thermal annealing, the mean sheet conductance of ambient devices improves by ~50% to 22.8±2.9 µS/□ but remains far below that of inert devices. These measurements indicate that while thermal annealing can remove weakly physisorbed molecules to boost device conductance, chemisorbed molecules cannot be removed, limiting device performance.

We have identified the detrimental effect of oxygen on 2D TMD devices that were previously thought to be ambient stable. Adopting an oxygen-free fabrication approach, we push our device performance toward the theoretical phonon-limited intrinsic mobility. Our work highlights that atomically thin materials require radical strategies beyond conventional nanofabrication. This provides guidance for the community, which should adopt practices to limit oxygen exposure to 2D materials. Our study advances the understanding of 2D TMDs that can likely be expanded to other atomically thin vdW systems. Thousands of 2D materials have been theoretically predicted since the first isolation of graphene,[68] but experimental research has been largely restricted to a handful of 2D materials that are more ambient stable. Our oxygen-free fabrication approach should accelerate the fundamental understanding of 2D materials and their translation for practical applications.

## Methods

**Materials:**

$WS_2$ and $MoS_2$ bulk crystals obtained from 6Carbon Technology and HQ Graphene were used for exfoliation. Commercial monolayers of $MoS_2$ and $WS_2$ grown by chemical vapor deposition (CVD) on sapphire substrates from 6Carbon Technology were employed for CVD-based devices. A 300 nm $SiO_2$-coated doped Si wafer served as the device substrate and gate dielectric in back-gated field-effect transistor (FET) device architectures.

**Sample preparation:**

All fabrication steps for inert devices were carried out inside a glovebox with $O_2$ and $H_2O$ levels maintained typically below 1 ppm. Initially, fresh substrates were cleaned under $N_2$ purging, followed by baking at 100 °C for 10 minutes. Exfoliation of 2D layers was performed using "blue-Nitto" low-adhesive tape (@Nitto Denko Corp.). First, thicker layers were cleaved from parent crystals and exfoliated 5-6 times within the same tape to thin down the layers and cover the entire tape surface. Subsequently, the 2D layers were lightly pressed onto the cleaned substrates immediately after removal from the hot plate. Finally, the samples were inspected under an optical microscope to identify the desired thickness of 2D layers.

CVD-grown monolayers were transferred from sapphire to the $SiO_2$/Si substrate using a polymer-assisted wet-transfer method. The CVD TMD films were delaminated from their growth substrates under ambient conditions and transferred inside the inert glovebox after vacuum cycling in the antechamber. We conducted hotplate annealing before the final film transfer to the device substrate for fabrication. These combined annealing and vacuum treatments aim to eliminate physisorbed adsorbates.

**Nanolithography:**

Contact patterning was conducted in an inert glovebox using the thermal scanning probe lithography (*t-SPL*) system (Nanofrazor Scholar, Heidelberg Instruments). A bilayer *t-SPL* resist consisting of 30/110 nm PPA/PMMA-MA was spin-coated, followed by baking at 140 °C for 90 s and 110 °C for 120 s, respectively. Details for the pattering parameters can be found in our previous study.[69] After thermal patterning, the exposed PMMA-MA layer was etched using 20 °C ethanol for 15 seconds. The In/Au metals (5/35 nm) were deposited by thermal evaporation at ~ 4 x $10^{-8}$ bar with a substrate temperature of 0 °C.

To rule out the effect of the nanolithography method used in device fabrication on the observed improvement, we also define the exact width of metal contact by electron-beam lithography (EBL). Samples were spin-coated with PMMA A5 electron resist at 4000 rpm for 90 seconds, followed by hard baking at 180 °C for 2 minutes before removal from the glovebox. The resist-coated 2D layers were subsequently transferred to the EBL system. Next, the contacts were written by EBL, and In/Au metal (5/30 nm) was deposited via thermal evaporation. All the intermediate processes, such as sample development in MIBK:IPA (1:3), metal lift-off using 1165 solvent (Microposit), and further resist coating for the next step of lithography, were performed inside gloveboxes using interconnected sample-transfer lines without exposing the 2D materials to air at any stage.

Finally, the electrical connections and bond pads were defined by EBL, followed by 5/55 nm Cr/Au metal evaporation.

**Ambient Environment Fabrication:**

All process steps and parameters, from exfoliation to resist coating, development, and liftoff, were explicitly performed in an air-atmosphere environment outside the glovebox.

**Electrical Characterizations:**

The electrical measurements were carried out using a cryostat probe station under a base pressure of ~$1\times10^{-4}$ mbar. The system was cooled down to 80 K using liquid nitrogen. Keithley 2450 source meters and a Lakeshore temperature controller were utilized for the measurements.

**Computational details:**

A monolayer TMD was considered with a single chalcogen vacancy as a defect site, and the influence of different defect densities was investigated. The amount of charge modulation in TMDs due to molecular adsorption is ascribed by $\Delta q$. The orientation of $O_2$ molecules determines the actual $E_{ad}$ and $\Delta q$ values. For example, side-on O-O configurations exhibiting stronger adsorption ($E_{ad}$ = 2.336 eV) and higher electron transfer ($\Delta q$ =1.080 e) compared to the end-on arrangements ($E_{ad}$= 1.98 eV and $\Delta q$ =1.031 e). The equilibrium band structure and atomic projected density of states (APDOS) are illustrated in ESI. Details about the DFT calculations and *ab initio* molecular dynamics (AIMD) simulations can be found in the Supporting Information.

**Acknowledgments**

We acknowledge the funding support from the Agency for Science, Technology and Research (A*STAR) (#21709, C230917006, and C230917007). C.S.L. acknowledges the support from A*STAR under its MTC IRG grant No. M23M6c0103 and MTC YIRG grant no. M21K3c0124. K.E.J.G. acknowledges support from a Singapore National Research Foundation Grant (CRP21-2018-0001). D.H. acknowledges funding support from A*STAR Project C222812022 and MTC YIRG M22K3c0105. Y.S.A. is supported by the Singapore Ministry of Education Academic Research Fund Tier 2 (Award No. MOE-T2EP50221-0019). Y.S.A. and S.W. are supported by the SUTD-ZJU IDEA Thematic Research Grant under the award number SUTD-ZJU (TR) 202203.

**Contributions**

S.M. and C.S.L. designed the experiment and carried out the transport measurements and result analysis. S.W. and Y.S.A. performed DFT and AIMD simulations. S.M., D.V., T.T.D, R.L. and Y.W.T contributed to the device fabrication and measurements. Y.T., S.M. and C.S.L. contributed to the python coding. F.B. and T.M. performed XPS studies. I.A.V., D.H., A.M., J.W.J, S.D. and K.E.J.G contributed to result interpretation and discussion. K.E.J.G., Y.S.A. and C.S.L. supervised the experiments and calculations. C.S.L. conceived the study. C.S.L. and S.M. wrote the manuscript with input from all authors.

**Competing Interests**

The authors declare no competing interests.

**Extended Data**

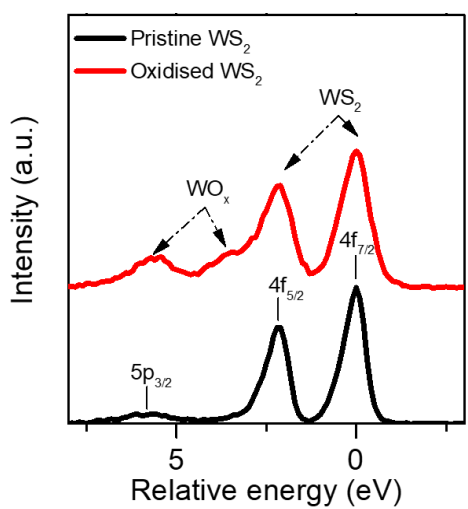

**Extended Data Figure 1:** X-ray photoemission spectroscopy of the W 4f and W 5p energy regions of pristine, freshly cleaved, and oxidized $WS_2$ samples. The data intensity (energy scale) is normalized (aligned) to the peak maximum intensity (position) for comparison purposes.

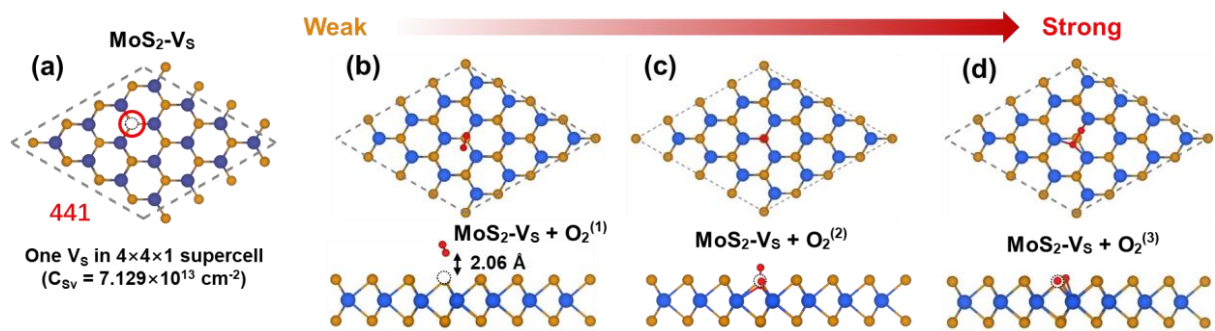

**Extended Data Figure 2:** Molecular adsorption on negatively charged $V_S$ sites on the TMDs' basal plane. (a) Single defect formation in a (4x4x1) supercell as a chalcogen vacancy ($V_S$), resulting in a defect density of ~7x10$^{13}$ cm$^{-2}$. Unlike the charged neutral state, with a single $V_S$ site, we find three possible, stable configurations for molecular $O_2$ adsorption based on their center of mass position relative to the $V_S$ site (defined as $O_2^{(1)}$, $O_2^{(2)}$, and $O_2^{(3)}$). These geometric configurations for $O_2$ adsorption and charge delocalization arise from the readjustment of TM-TM bonds relative to the $V_S$ site, as described by the Jahn-Teller distortion mechanism. Notably, if the distance between an $O_2$ molecule and the $V_S$ site is considerably larger (h~2.06 Å), only physisorption occurs due to weak interactions (as shown in (b)). However, strong chemisorption is observed when $O_2$ occupies the $V_S$ site, leading to significant charge transfer and *p*-doping. Two possible chemisorption configurations can occur depending on the lowest energy states: (c) vertical and (d) horizontal with respect to the 2D basal plane.

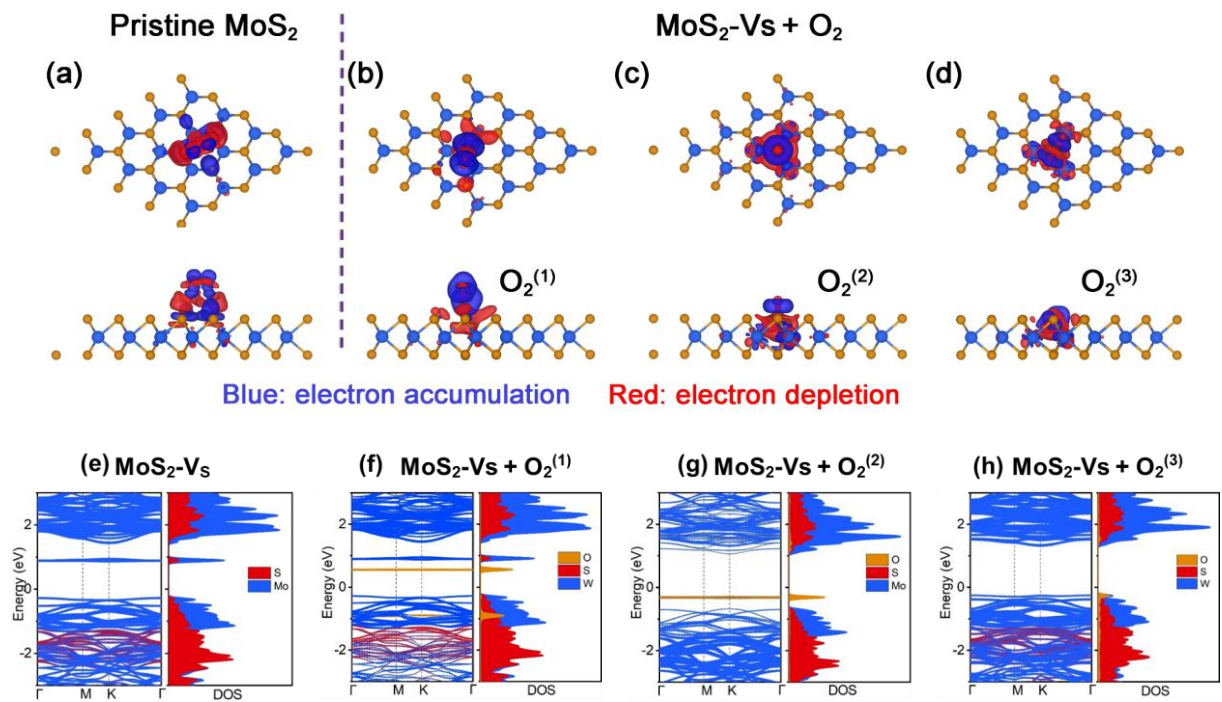

**Extended Data Figure 3:** Charge density difference after $O_2$ molecule adsorption on (a) pristine $MoS_2$, and (b-d) defective $MoS_2$ sites. The energetically favorable arrangements for (b) physisorption and (c & d) chemisorption are shown here. Band structure and atomic projected density of states (APDOS) for (e) single chalcogen vacancy ($V_S$) in $MoS_2$ basal plane, and $MoS_2$ with adsorbed molecules, such as (f) physiosorbed $O_2$, and (g-h) chemisorbed $O_2$.

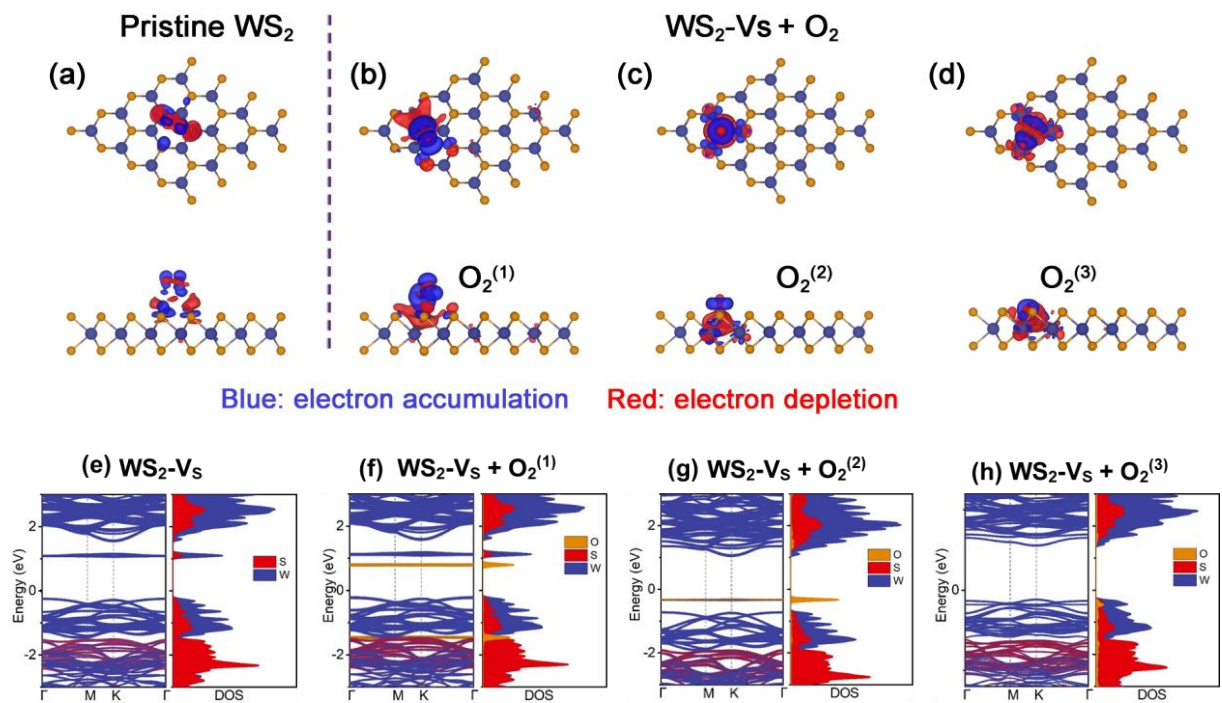

**Extended Data Figure 4:** Charge density difference after $O_2$ molecule adsorption on defective $WS_2$ sites. The energetically favorable arrangements for (b) physisorption and (c & d) chemisorption are shown here. Band structure and atomic projected density of states (APDOS) for (e) single chalcogen vacancy ($V_S$) in $WS_2$ basal plane, and $WS_2$ with adsorbed molecules, such as (f) physiosorbed $O_2$, and (g-h) chemisorbed $O_2$.

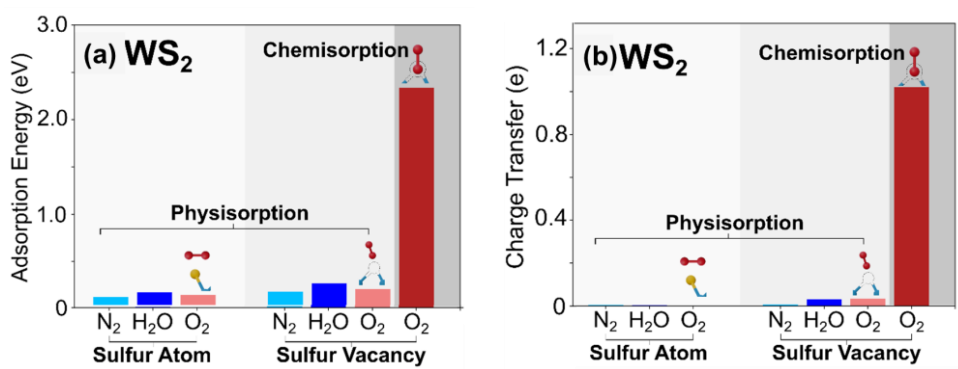

**Extended Data Figure 5:** Effect of oxygen molecular adsorption on monolayer WS$_2$. (a) Adsorption energy of O$_2$, N$_2$, and H$_2$O on the pristine surface and sulfur vacancy sites. (b) Estimated electron transfer from WS$_2$ to adsorbate molecules.

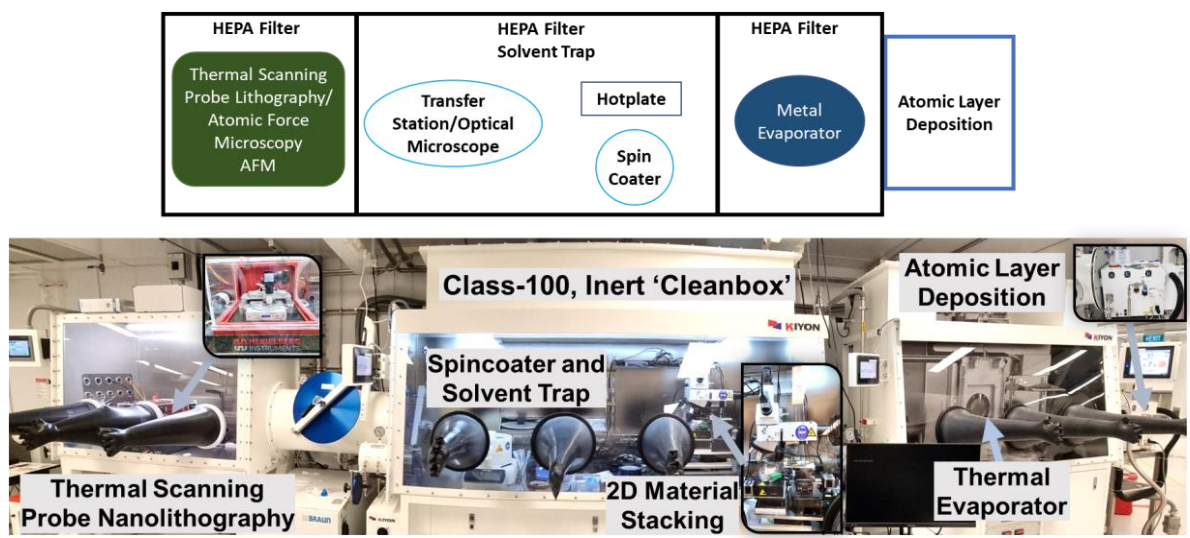

**Extended Data Figure 6:** Our purpose-built glovebox fabrication facility. This customized setup allows complete fabrication of 2D devices in an argon environment with typical O$_2$ and H$_2$O levels maintained below < 5 ppm, usually < 1 ppm. Capabilities include nanolithography, wet chemistry, 2D material transfer, metal deposition, and atomic layer deposition.

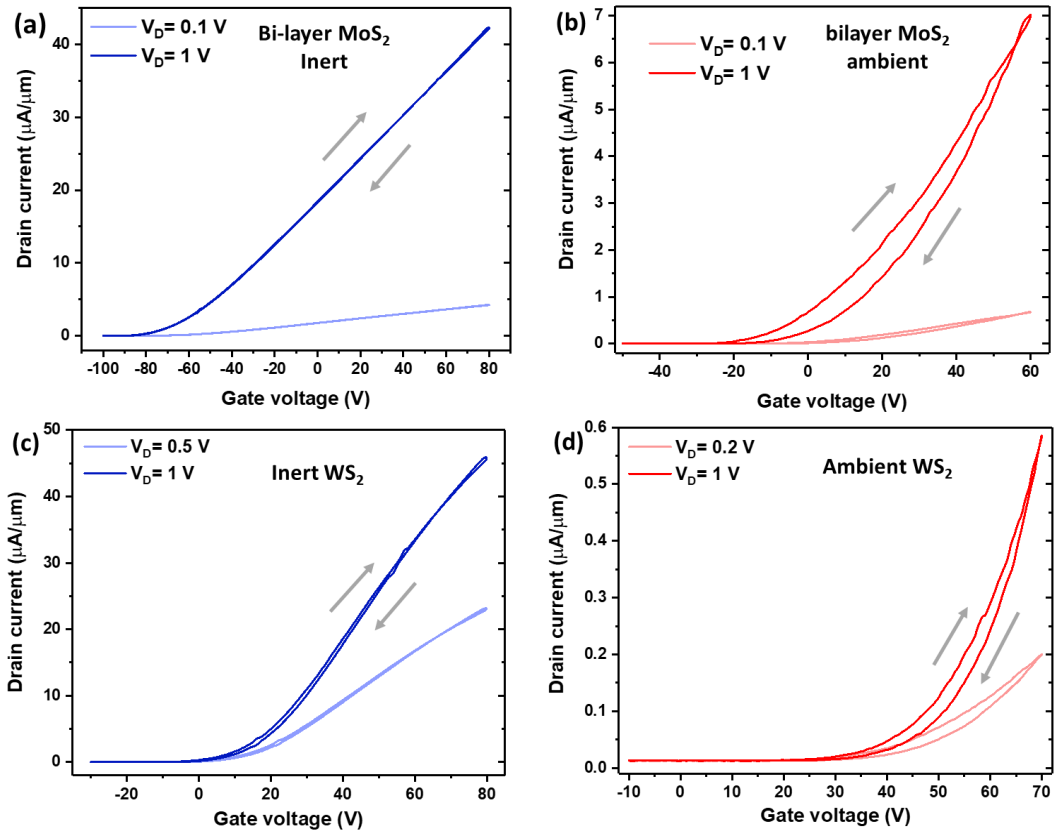

**Extended Data Figure 7:** Dual sweep transfer characteristics of inert and ambient (a-b) bilayer $MoS_2$ and (c-d) $WS_2$ FETs.

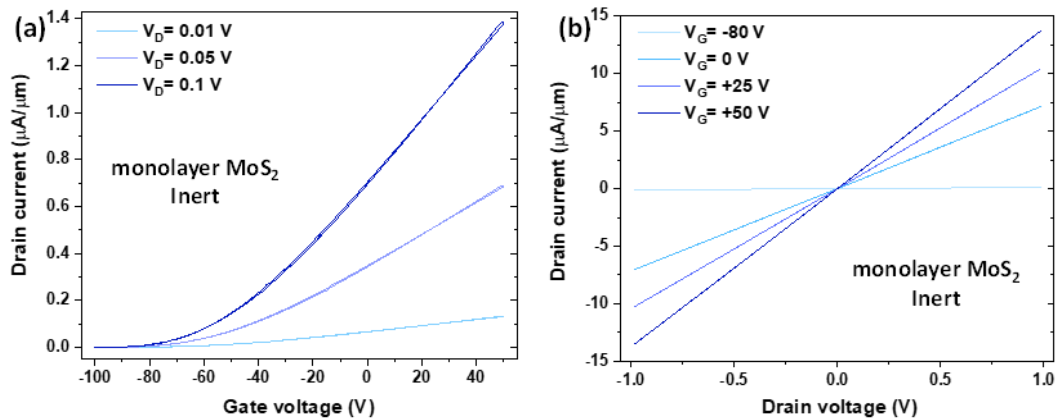

**Extended Data Figure 8:** Electrical characteristics of a typical inert monolayer $MoS_2$ FET. The transfer curves (a) and output curves (b) are presented for various drain and gate biases, respectively.

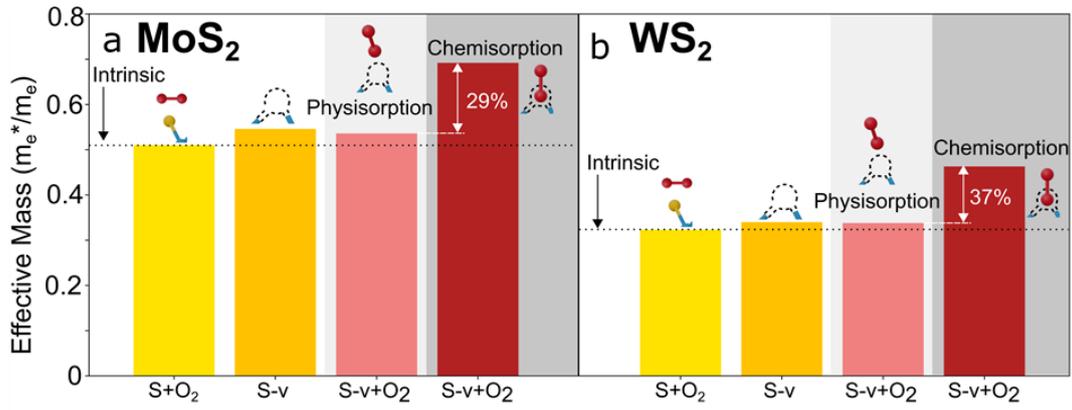

**Extended Data Figure 9:** The electron's effective masses in defective monolayers (a) $MoS_2$ and (b) $WS_2$, with various interaction configurations of $O_2$ molecules. The dotted line presents a visual guide for the change in effective mass compared to the pristine case.

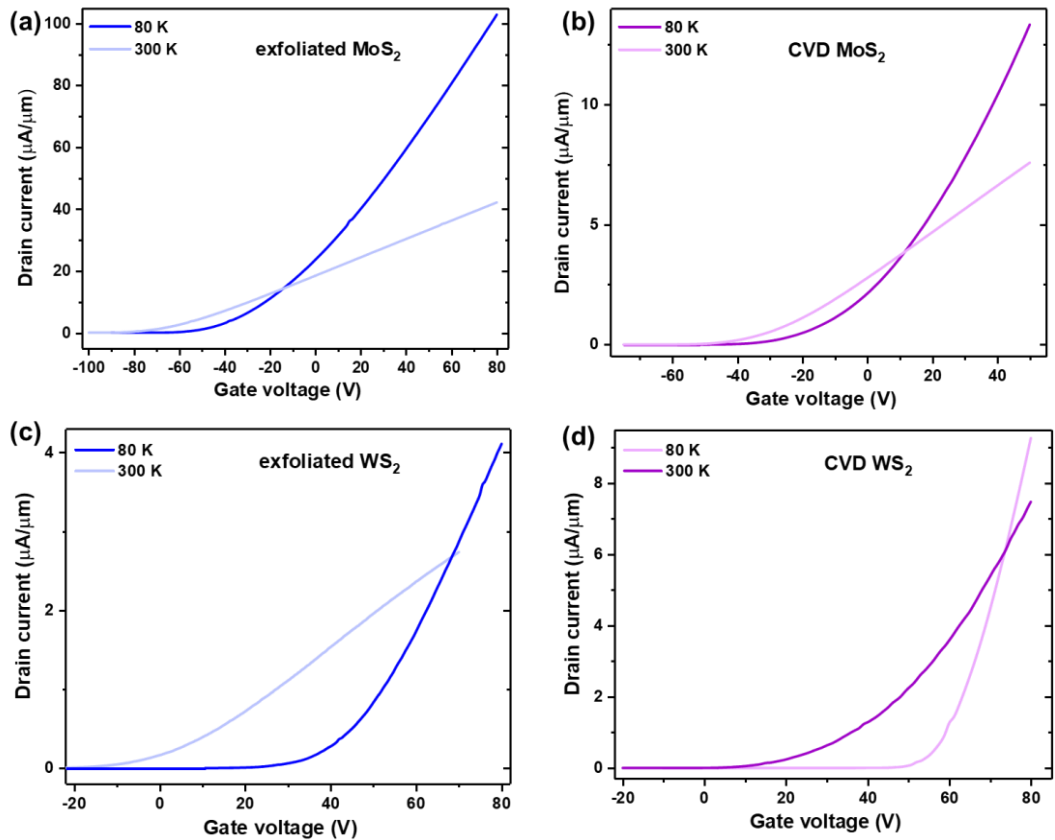

**Extended Data Figure 10: Transfer characteristics** at 300 K and 80 K for inert FETs. (a) exfoliated $MoS_2$, (b) CVD $MoS_2$, (c) exfoliated $WS_2$ and (d) CVD $WS_2$ FETs. Mobility is calculated for each case, and the summarized plot is presented in Figure 3c of the main text.

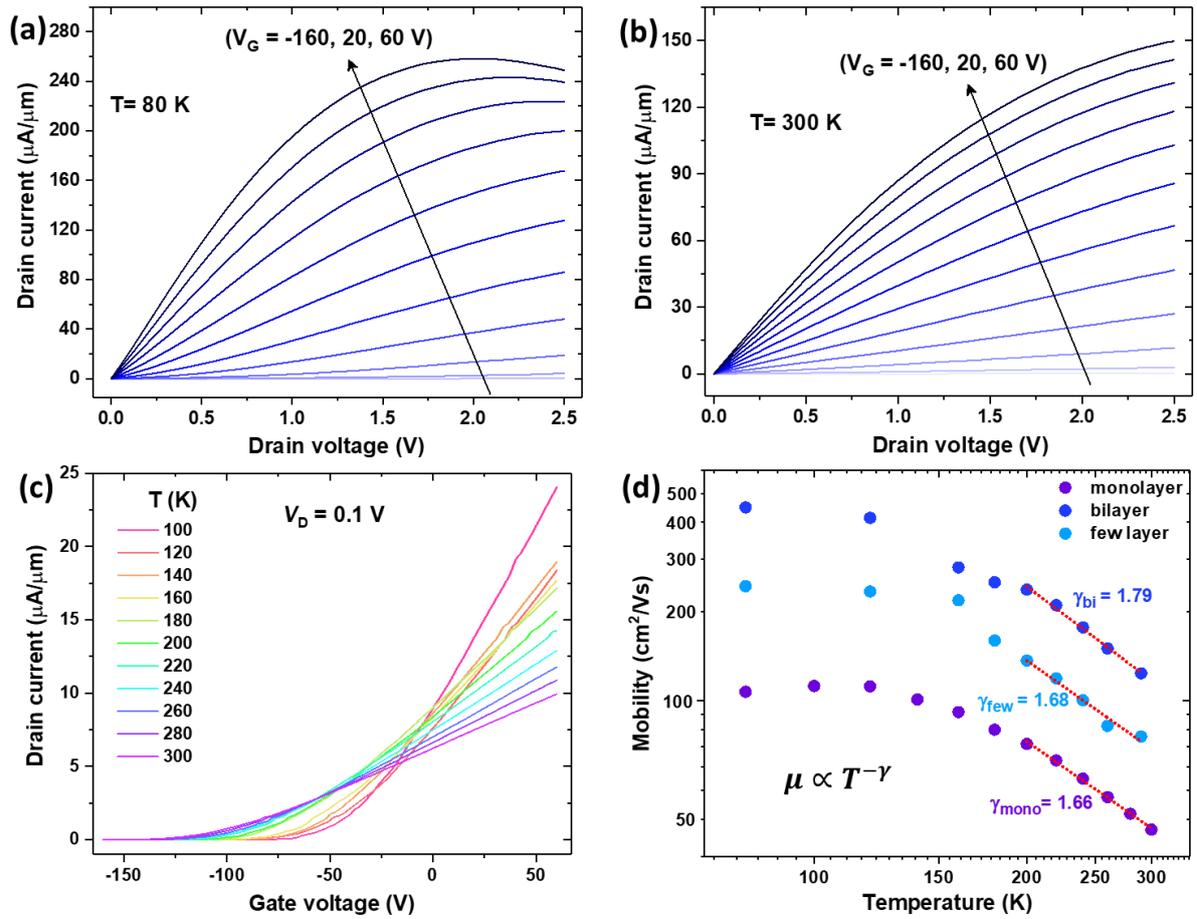

**Extended Data Figure 11:** Electrical data of inert MoS$_2$ FETs. Output characteristics of a monolayer device measured at (a) 80 K and (b) 300 K. (c) Temperature-dependent transfer characteristics of a monolayer device showing mobility improvement with decreasing temperature due to quenching of phonon scattering. (d) Mobility as a function of temperature ($\mu \propto T^{-\gamma}$) showing phonon-limited behavior at higher temperatures.

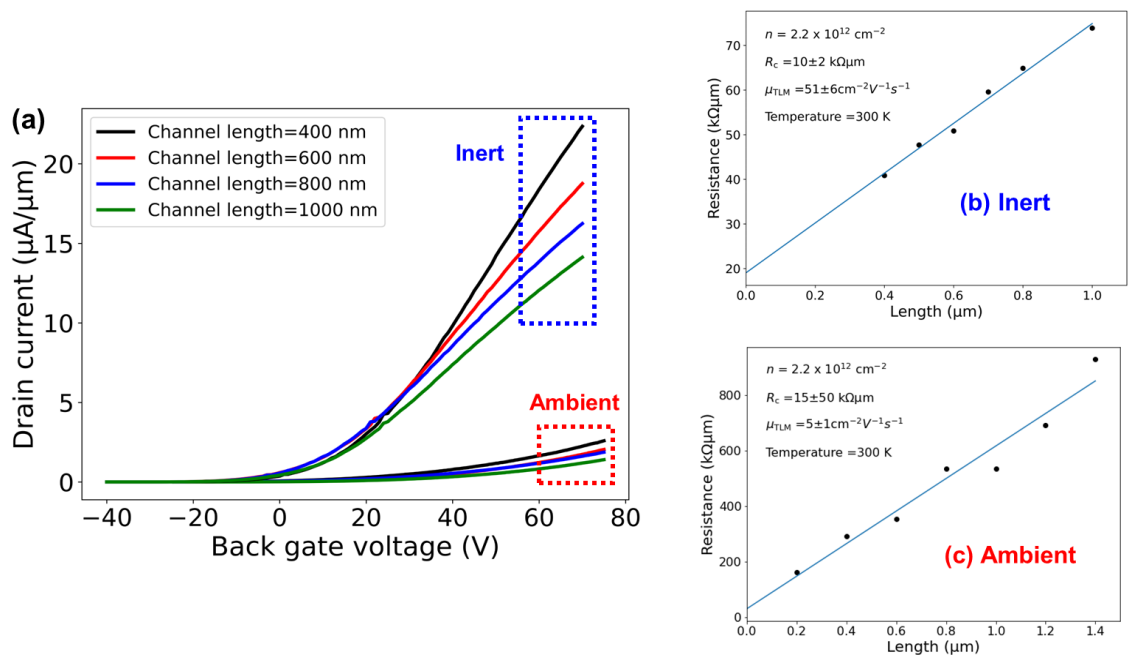

**Extended Data Figure 12:** (a) Transfer characteristics comparing inert and ambient few-layer $MoS_2$ transistors of different channel lengths in a transfer length method (TLM) geometry measured at $V_D$ = 1 V. Extraction of TLM contact resistance ($R_C$) for (b) inert and (c) ambient devices at a carrier density of $2.2 \times 10^{12}$ cm$^{-2}$. Inert devices show ~10× improvement in the TLM mobility and lower contact resistance $R_C$.

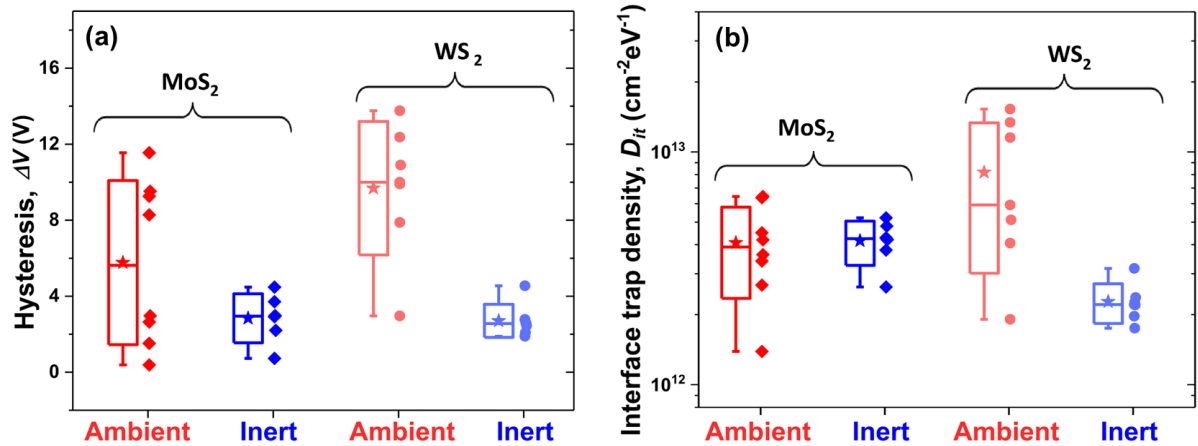

**Extended Data Figure 13:** (a) Boxplot showing the distribution of threshold voltage hysteresis determined for inert and ambient devices at room temperature. The back gate voltage $V_G$ is swept from -30 to +70 V (forward sweep) and then back to -30 V (backward sweep) with a constant $V_D$ = 1 V. The hysteresis is then calculated from the difference of the threshold voltages determined from the forward and backward sweeps. (c) Boxplot showing the distribution of the interface trap density of inert and ambient devices calculated from the subthreshold slopes.

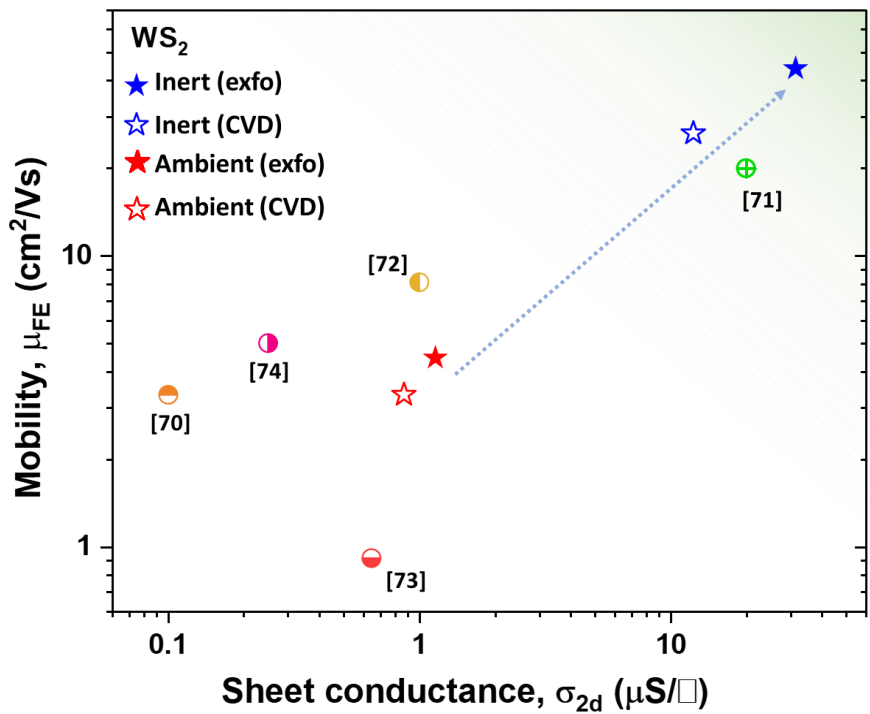

**Extended Data Figure 14:** Benchmarking FET performance against the existing literature for exfoliated and CVD $WS_2$. Solid (hollow) stars represent exfoliated (CVD) FETs.[70–74]

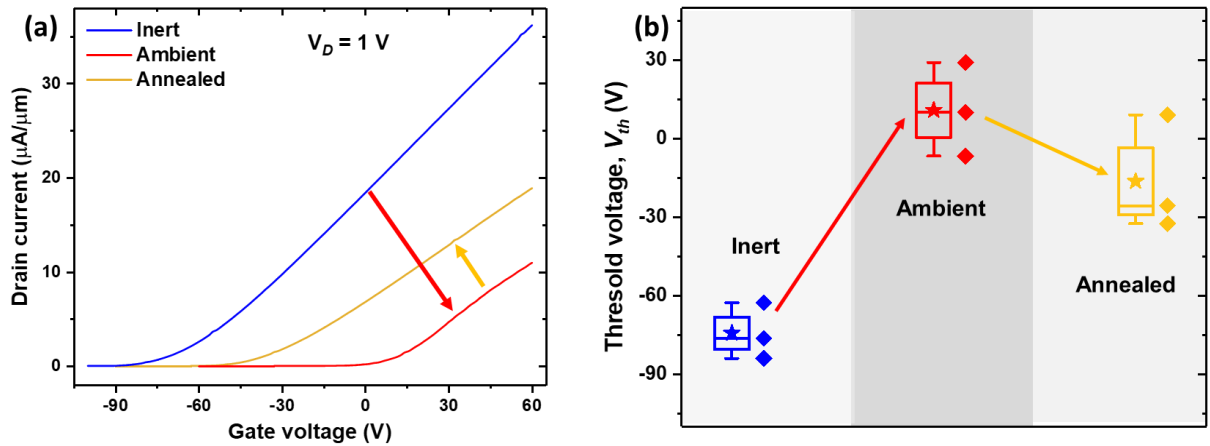

**Extended Data Figure 15:** (a) Transfer characteristics and (b) threshold voltages of inert and ambient devices. Ambient devices are subsequently annealed in a 10% Ar/H$_2$ forming gas at 200 °C for 2 h. While annealing can partially restore the current and threshold voltage, complete reversal is not possible.

Transistors Realized by Multilayer Graphene Electrodes and Application to High Responsivity Flexible Photodetectors. *Adv. Funct. Mater.* **27**, 1703448 (2017) 10.1002/adfm.201703448.